\documentclass{aa}  

\usepackage{graphicx}
\usepackage{placeins}
\usepackage{txfonts}
\usepackage{hyperref}
\usepackage{mathrsfs} 
\usepackage{soul}
\usepackage{siunitx}

\usepackage{natbib}
\bibpunct{(}{)}{;}{a}{}{,}

\usepackage{color} 
\usepackage{amssymb}

\usepackage{caption}
\usepackage{subcaption}

\usepackage[normalem]{ulem}

\newcommand{\ndof}{n_{\rm dof}}
\newcommand{\sfn}{\sigma_{\rm fn}}

\begin{document} 

    \title{Evidence of the association of repeating fast-radio-burst sources with fast-spinning super-twisted magnetars}

\titlerunning{FRB from super-twisted magnetars}

   \author{G. Voisin\inst{1}\fnmsep\thanks{Email: guillaume.voisin@obspm.fr}
          \and
          F. Francez\inst{1} 
         }

\institute{LUX, Observatoire de Paris, Universit\'e PSL, Sorbonne Universit\'e, CNRS 92190 Meudon, France
    }
   
   \date{Received XXXX XX, 2020; accepted XXXX XX, 2020}

  \abstract
   {Fast radio bursts (FRBs) are bright millisecond radio events of unknown extragalactic origin. Magnetars are among the main contenders. Some sources, the repeaters, produce multiple events but so far generally without the characteristic periodicity that one could associate with the spin of a neutron star. }
    {Assuming that the bursts originate from a magnetar magnetosphere, we aim to fit our geometrical model to the two main repeaters of the CHIME/FRB catalogue, namely FRB 20180814A and FRB 20180916B, and thus characterise the star.}
    {The model can generate dynamic spectra that can be directly compared to FRBs. We applied nested sampling in order to evaluate the main parameters of the model. These parameters being common to all bursts from a given repeater, they were fitted together as a single dataset.}
    {We constrained the spin and magnetic parameters of the star, which were encoded into burst spectro-temporal morphologies. We estimate that a very strong toroidal magnetic component together with spin periods of, respectively, $2.3_{-0.5}^{+0.5} ~ \rm s$ and $0.8_{-0.2}^{+0.1} ~ \rm s$ best explain the data. We argue that this points towards young magnetars with super-twisted magnetospheres, and possibly low-field magnetars.
     }
   {}


   \maketitle

\section{Introduction}\label{sec:intro}
Fast radio bursts (FRBs) are extremely bright radio pulses that typically occur within milliseconds. Their dispersion measure (DM) -- the spectro-temporal pattern imprinted in the signal by the interstellar medium -- can be used to infer a distant extragalactic origin, which has been confirmed by the localisation of some host galaxies \citep[e.g.][]{petroff_fast_2022}. 

Two broad categories stand out \citep[e.g.][]{pleunis_fast_2021}: first, the one-off events that have been seen only a single time; second, the so-called repeaters that have been seen to repeat in a random, albeit clustered, fashion \citep[e.g.][]{cruces_repeating_2021}. Two repeaters have been shown to follow a periodic activity window \citep{rajwade_possible_2020, collaboration_periodic_2020}. 

Spectro-temporal burst morphologies appear to be statistically different between the two categories \citep[e.g.][]{pleunis_fast_2021}. One-offs appear to be shorter and broad-band, and usually made of a single component, or sub-burst. Repeaters, on the other hand, are narrow-band events that tend to last longer, with successive components drifting downwards in frequency \citep{hessels_frb_2019}. This difference in morphology leads one to wonder if indeed there might be two different types of progenitors. 

Many theories have been proposed to explain FRBs, an increasing number of which are focusing on neutron stars, and particularly variants of magnetars \citep[e.g.][]{zhang_physical_2020, voisin_maze_2021}. Magnetars are appealing because of their ample magnetic energy reservoir, and their known randomly eruptive behaviour, which appears to be statistically consistent with FRBs \citep{wadiasingh_repeating_2019} although it occurs primarily in the X-ray and gamma-ray bands with occasional radio counterparts. These radio counterparts are, however, much weaker than known FRBs and, conversely, high-energy counterparts have so far not been detected at extragalactic distances \citep{zhang_multiwavelength_2024}. Thus, if FRBs are caused by magnetars, then these are likely unusual by Galactic standards. One putative FRB has been seen from the Galactic magnetar SGR1935+2154 \citep{bochenek_fast_2020, andersen_bright_2020}, with a burst both a thousand times brighter than other known magnetar emissions, and a hundred times weaker than extragalactic FRBs. 
A range of models involve magnetars, including FRBs, which are caused by very young magnetars \citep{metzger_millisecond_2017} or low-twist ones \citep{wadiasingh_repeating_2019}.

Inferring source properties from observed FRBs proves to be very challenging. So far, to the best of our knowledge, events have been fitted independently using empirical functions \citep[e.g.][]{hessels_frb_2019, fonseca_modeling_2024}. This allows one to extract burst-specific quantities such as the bandwidth, duration, or frequency drift rate. 

In \cite{voisin_geometrical_2023}, we proposed a geometrical model that can constrain the spectro-temporal properties of an event as a function of its time of arrival and source parameters. The model assumes that emission is due to a localised ultra-relativistic plasma flowing along curved streamlines. Tangential emission along magnetic field lines is notably at the core of the rotating vector model \citep{radhakrishnan_magnetic_1969, petri_polarized_2017} successfully applied to explain the polarisation swing of many pulsars. Similar swings have been observed in a couple of FRBs \citep{luo_diverse_2020, mckinven_pulsar-like_2025}. This suggests that this type of geometrical model is also relevant for FRBs, albeit with possibly different magnetic configurations or lines of sight.

In our model the emission mechanism need not be specified, and it is enough to say that emission is beamed forwards as a result of relativistic beaming. The model shows that the spectro-temporal morphology of a burst is constrained by a geometrical envelope that depends only on the local geometry of the streamlines, the spin period of the star, and the degree of beaming. Thus, the morphology of events from a given source are constrained by its spin and global magnetic geometry. Said differently, global source properties can be inferred from burst morphologies.

In this work, we assume that FRBs are produced by plasma flowing along magnetic field lines inside the co-rotating region of a magnetar magnetosphere. Such magnetospheres are thought to be twisted; that is, to possess a toroidal magnetic component. This can result from magnetic footpoint motion during eruptions \citep{thompson_electrodynamics_2002}, or from the formation and structure of the star \citep[e.g.][]{barrere_new_2022, uryu_equilibriums_2023}. We used a very simplified phenomenological model of the magnetic structure, considering it as the sum of a dipole aligned with the spin axis, and of a uniform toroidal component decaying radially as a power law (see Sec. \ref{sec:methods}).

Using our model, we performed a fit of the data of the two most frequent repeaters in the first CHIME/FRB catalogue \citep{collaboration_first_2021}: FRB20180814A and FRB20180916B, which in the following we call A and B, respectively. B is famous for being one of the two repeaters exhibiting a periodic activity window \citep{collaboration_periodic_2020}.

All the events of a given source are fitted jointly, since the model predicts common morphological constraints depending mostly on spin and magnetic geometry. This is also why repeaters are more relevant than one-off events for the practical tests of our model. To our knowledge, it is the first time such a joint fit has been performed. 
Although the number of bursts is relatively low, the fit has the advantage of keeping computational needs manageable for this exploratory work.

\section{Methods}\label{sec:methods}
\subsection{Data} \label{sec:data}

The data for the two sources with the largest number of repetitions in the CHIME/FRB catalogue \citep{collaboration_first_2021}, namely FRB 20180814A (A) and FRB 20180916B (B), was retrieved from the dedicated website\footnote{\url{https://chime-frb-open-data.github.io}}. The number of events is, respectively, 11 for A and 19 for B. The catalogue itself was retrieved thanks to the \texttt{python} package \texttt{cfod} \footnote{\url{https://github.com/chime-frb-open-data/chime-frb-open-data?tab=readme-ov-file}}.

The data spans from August 2018 to June 2019 for A, and from September 2018 to June 2019 for B. In total, the public data in the catalogue contains 11 events for A and 19 events for B as determined by the CHIME/FRB pipeline. Events correspond to one or several components forming a continuous signal, with the notable exception of FRB20181019A from source B for which the two components appear to be fully separated by about $60 \;$ms. In the case of A, seven out of eleven events are composed of a single component (also called sub-burst below) and the four others have multiple components. For B, there are eleven singles and eight multiples. Both sources have events across the entire bandwidth of 400-800 MHz, with durations shorter than 100 ms. In the catalogue, each event is de-dispersed individually, with DMs of $\sim 190 \; \rm pc\cdot cm^{-3}$ for A and $\sim 349 \; \rm pc\cdot cm^{-3}$ for B (see also Sec. \ref{sec:priors}).

Basic pre-processing such as interference removal and generation of dynamic spectra (waterfall plots) was adapted from the documentation kindly provided by the CHIME/FRB collaboration\footnote{In particular \url{https://chime-frb-open-data.github.io/waterfall/}.}. The original time resolution of 0.983 ms was kept, while the frequency resolution was reduced to 128 bins corresponding to bin widths of 3.125 MHz in order to reduce the computational load. 

For each event, quantities provided in the catalogue, such as the burst arrival time, width, and fluxes, were retrieved in order to be used as first guesses for the fitting procedure. Gaussian profiles were then fitted to the intensity profile, requiring manual interventions in a select number of cases in which many components were needed. These fits finally provided initial estimates for the peak times, Gaussian width, and fluence of each individual component within events.

\subsection{Model} \label{sec:model}

The model presented in \citet{voisin_geometrical_2023} constrains burst spectro-temporal properties using the geometry of streamlines followed by the emitting plasma. Its key assumptions are i) the relativistic nature of the emitting plasma and ii) the locality of the emission region. The main consequence of assumption i) is that the emission is relativistically beamed forwards in the direction of motion. Assumption ii) ensures that streamlines within the emission region can be approximated to a single streamline endowed with an effective emission angle. Thus, this emission angle accounts for the local field line divergence as well as the intrinsic beaming of the emission process. Locally, streamlines are approximated by their tangent direction and curvature. The fact that a line is curved together with the narrowness of the emission cone ensures that only a small segment of the line is visible by an observer at any time. The small lateral extent of the emission region effectively means that only a small bundle of lines radiates. In addition, the model accounts for streamlines defined in a rotating frame with period $P_*$. 

\begin{figure}
	\centering
	\includegraphics[width=0.46\textwidth]{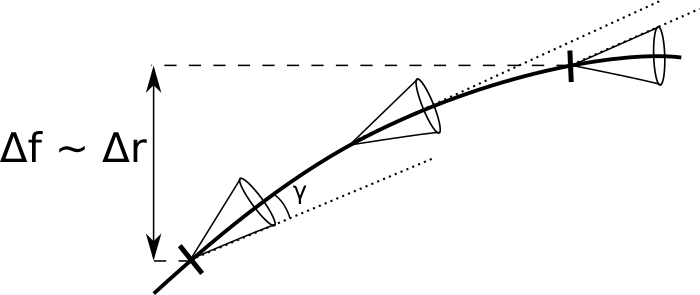}
	\caption{Illustration of the model, from \citet{voisin_geometrical_2023}. The emitter travels from left to right, emitting in a forward cone (the cone is illustrative, in practice we use a smooth angular profile in this work). Emission frequency varies by $\Delta f$ as it travels radially by $\Delta r$. Dotted lines show the observer's direction, with $\gamma$ the angle with respect to the path. The two thick ticks delimit the visible segment beyond which no emission can be seen from within the cone. The characteristic emission frequency, $f_c$, maps to the radius, $r_c$, at the centre of the segment. The visible segment being in a rotating frame, it varies with time until it vanishes.}
	\label{fig:sketchr2freq}
\end{figure}

In this work, we assume that streamlines follow the magnetic field lines of a co-rotating neutron-star magnetosphere. This implies that emission regions are located sufficiently close to the star in order to avoid relativistic co-rotation speeds that are not accounted for by the model as yet. This condition was not imposed a priori, but we checked a posteriori that this is indeed the case that $r  << cP_*/2\pi$, where $r$ is the distance to the stellar centre, $c$ is the speed of light, and the right-hand side is the so-called light-cylinder radius, which is the distance at which the co-rotation speed equals the speed of light. 

The emission frequency, $f$, was mapped to geometry through the radius-to-frequency mapping phenomenological law \citep[e.g.][]{lyutikov_radius--frequency_2020}, $f = f_* R_*/r$, where $R_*$ is the stellar radius and $f_*$ is the emission frequency at the stellar surface. Thanks to this mapping, the emission region maps directly onto an area in the dynamic spectrum of the observer. Indeed it is clear that, as the star rotates, the observer moves out of the emission beam, and thus defines a finite visible segment of time, while the radial extent of the emission region maps to a frequency interval (see Fig. \ref{fig:sketchr2freq}). The radial extent of the region  within the line of sight of the observer changes in time, which maps to a non-trivial spectro-temporal envelope\footnote{See also the animation at \url{https://astrotube.obspm.fr/w/f9SxoEzzc2bBvc5KjXAKQv}}. The centre of each spectro-temporal envelope $(T_0, f_0)$ was mapped to a point in the magnetosphere. This point was numerically computed as the point at which the tangent to the field line points towards the observer at time $T_0$ at the altitude corresponding to $f_0$.

\newcommand{\vB}{\vec{B}}
The magnetic field was constructed in the simplest possible way: a dipolar field, $\vB_{\rm d}$, superposed with a toroidal field (relative to the rotation axis) defined by $\vB_{\rm t} = -\alpha B_{\rm d}{}_{*} (R_*/r)^p \vec{e}_{\phi}$, where $B_{\rm d}{}_{*}$ is the dipolar field strength at the pole, and $\vec{e}_{\phi}$ is the azimuthal unit vector. Magnetic geometry is thus parametrised by the toroidal-to-dipolar ratio, $\alpha$, and by the exponent $p$. The ratio, $\alpha$, can be either positive or negative. The strength, $B_{\rm d}{}_{*}$, is irrelevant for geometrical purposes and can be set to 1. In order to simplify further, the dipole is assumed to be aligned with the spin axis of the star, making the geometry cylindrically symmetric. It follows that the orientation of the star with respect to the observer is only determined by the inclination, $i$, of its axis with respect to the normal to the plane of the sky (oriented away from the observer), and the rotational phase is irrelevant in this particular case. If the toroidal field is important, or dominant, axial symmetry is partially realised, which makes it a more reasonable approximation. 

We also tried a model where $f = f_* R_*^p/ r^p$, such that the radius-to-frequency mapping follows the same scaling as the toroidal component of the magnetic field (which, in our model, is expected to dominate in the inner magnetosphere). However we were unable to draw conclusions from this model, as the MCMC could not converge and was mostly constrained by priors on magnetic parameters and frequency parameters. Thus we focused only on the inverse law mentioned above. 

So far, we have reasoned with a finite emission cone emerging from a radiating moving point. In practice, bursts were modelled using a pseudo-Gaussian model \citep{voisin_geometrical_2023} whereby a radiating blob (it could be a wave packet or an actual blob of matter) is injected along a field line following a Gaussian time profile in intensity with a characteristic width, $w$, fluence, $F$, and injection time, $t_i$. A blob radiates across a Gaussian angular profile of characteristic opening angle $\Omega$. This angle is assumed to be common to all events of a given source, although this is probably not the case in general.

Under the pseudo-Gaussian model (with $f\propto r^{-1}$), an event is the sum of sub-bursts of the form (see Eq. (63) of \citet{voisin_geometrical_2023})
	\begin{equation}
	I(t_a,f) \propto  \underbrace{\frac{1}{f^2\Omega^2\cos\rho} \exp\left(-\frac{\gamma^2(t_a, f)}{2\Omega^2}\right)}_{{\cal E}(t_a,f)} F \exp\left(-\frac{(t_a-t_i)^2}{2w^2}\right),
	\end{equation}
	where $I$ is the intensity per unit time of arrival, $t_a$, per unit frequency, $f$. The function $\gamma$ characterises the emission beam, and is defined as the angle between the line of sight and the tangent to the magnetic field at the magnetospheric location that produces the emission received at $(t_a,f)$ in the dynamic spectrum (see Fig. \ref{fig:sketchr2freq}). This function depends on the global geometry, orientation, and spin; that is, it depends on the global parameters. It also depends on the location of the centre of the emission region, $\vec{x}_0$, which is given by the co-ordinates $(T_0, f_0)$, it being understood that the model maps time and frequency to the magnetospheric location. The angle $\rho$ is the angle of the magnetic field with the radial direction at $\vec{x}_0$. 
	
	As a result, each sub-burst defined by a triplet $(F, t_i, w)$ is a Gaussian modulated by what may be called the envelope function, $\cal E$, of the emission region at $x_0$. Importantly, it is independent of the sub-burst parameters $(F, t_i, w)$ but rather delimits where in the dynamic spectrum sub-bursts can be seen whenever they occur. The envelope only depends on the global parameters and the location of the emission region. It is thus the key element that links all sub-bursts of all events to the properties of the source.
	
\begin{figure}
	\centering
	\includegraphics[width=1\linewidth]{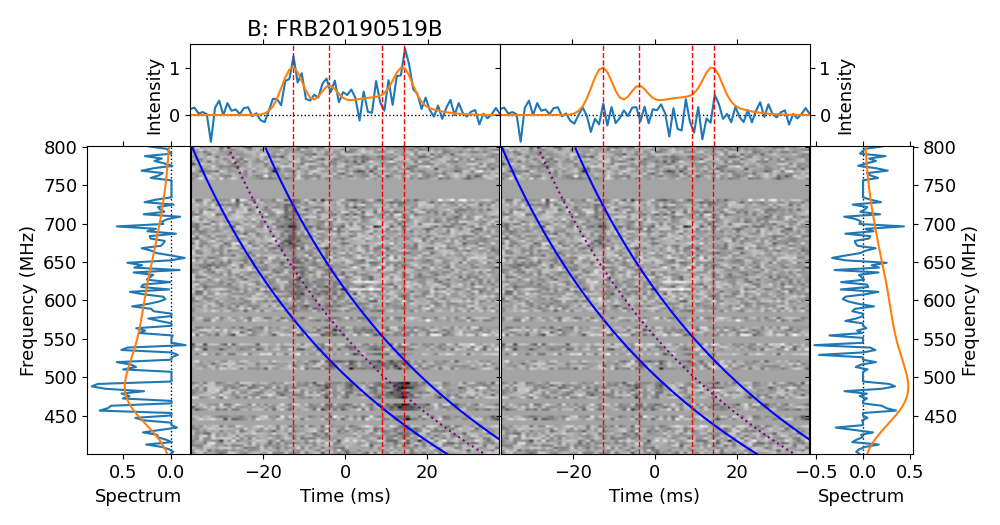}
	\caption{Residuals of a particular event from B, FRB20190519B, using best-fit parameters. Left-hand side: CHIME/FRB dynamic spectrum (middle), intensity curve (top), and spectrum (left). The one-$\Omega$ envelope (or emission angle iso-contour) of the emission beam is shown in blue, with a dashed purple line showing its characteristic frequency as a function of time. The vertical dashed red lines mark the injection time for each sub-burst. In the spectrum and intensity plots the blue lines represent the data, and the orange lines the model. 
			Right-hand side: Same but showing the residuals between the data and the model.
	}
	\label{fig:residuals}
\end{figure}

	The one-$\Omega$ envelope is the contour in the dynamic spectrum defined by $\gamma(t_a, f) = \Omega$. It is represented by a blue line in the residual plots of Fig. \ref{fig:residuals} and in the additional material available on the \href{https://dx.doi.org/10.5281/zenodo.16421753}{Zenodo repository}. It is important to keep in mind that sub-bursts can occur outside of this contour, both in time and frequency, indicating that the observer is seeing the outer part of the beam. 
	 We could also have used an iso-contour of ${\cal E}(t_a, f)$, which has an additional dependency on $f$, but using $\gamma$ presents the advantage of having an analytical expression \citep{voisin_geometrical_2023}, while the exponential makes it a good approximation.

In this work, we assume that each event in the CHIME/FRB catalogue is caused by a single emission region. This means that a single envelope function is associated with each event such that the two words can be used interchangeably. One could, however, imagine an event with a subset of sub-bursts associated with one envelope, and another subset associated with another envelope. In other words, emission would come from several regions simultaneously.

Propagation effects, the DM, and scattering are taken into account. In the CHIME/FRB catalogue, each event is de-dispersed with a specific DM and fitted for a specific scattering time. In this work, we fit only one DM value common to all events. It follows that we need to disperse the modelled bursts by the difference between the fitted DM and the CHIME/FRB DM. We fitted for a single common DM because it is the simplest approach, but also because our model predicts intrinsic effects that can mimic local DM variations \citep{voisin_geometrical_2023} and that might otherwise be difficult to disentangle from local DM variations. 
Pseudo-Gaussian bursts have a Gaussian temporal profile at a given frequency (but the parameters of the Gaussian vary in frequency). This is because the $\gamma^2(t_a, f)$ can be expanded to second order in $t_a$. As a result, the effect of scattering can be computed using the formula in \citet{mckinnon_analytical_2014}. In order to reduce the complexity of the fit, we kept the values provided by the CHIME/FRB catalogue fixed.

Altogether our model has seven global parameters $\left(f_*, i, P_*, \alpha, p, \Omega, \mathrm{DM}\right)$; that is, parameters that are common to all events of a given source. On the other hand, we have a relatively large number of local parameters: the two parameters $(T_0, f_0)$ for each envelope, and within it the triplet $(F, t_i, w)$ for each sub-burst  (see also Fig. \ref{fig:diagramfitting}).

Nonetheless, the predictive power of this model is superior to an empirical model that fits each sub-burst independently from one another, in the sense that the number of parameters required is smaller. In order to demonstrate that assertion, we compare the number of parameters of our model with a fully empirical model whereby each sub-burst is modelled with a two-dimensional Gaussian of the form $A\exp\left({}^{\rm T}(\vec{y} - \vec{m})C(\vec{y} - \vec{m})\right)$, where $\vec{y} = (t_a,f)$.
Each sub-burst then requires six independent parameters: one amplitude, $A$, two means, $\vec{m}$, and three parameters for the correlation matrix, $C$. Thus, one has $6 \bar n_{\rm b/e} n_{\rm e}$ parameters in the empirical model, where $n_{\rm e}$ is the number of events and $\bar n_{b/e}$ is the average number of sub-burst per event.
On the other hand, our model has $(2 + 3\bar n_{\rm b/e} ) n_{\rm e} + 7$  parameters: seven global parameters, two parameters $(T_0, f_0)$ per envelope (equivalent to events here), and three parameters $(F, t_i, w)$ per sub-burst. Thus, it is enough to have $n_{\rm e} > 7$ in order for our model to require fewer parameters than the empirical model, and a single event is sufficient if it has at least three sub-bursts. 

However, fitting such a large parameter space remains challenging. Since the total fluence of an event can be measured independently, we further reduced the parameter space by fixing the sum of the sub-burst fluences of a given event to be equal to the measured total fluence of that event. As a result, only $n_{\rm b/e} -1$ fluence parameters, $F$, were fitted per event (instead of $n_{\rm b/e}$). 

For A, $n_{\rm e} =11, \bar n_{\rm b/e}=22/11\simeq2$ such that we have 84 parameters (77 local ones), compared to 123 for an empirical Gaussian model (assuming total fluences to be fixed in both cases). 
For B, $n_{\rm e} =19, \bar n_{\rm b/e}=34/19\simeq1.8$ such that we have 128 parameters (121 local ones), compared to 185 for an empirical Gaussian model. 

We developed an implementation of the model as a \texttt{python} package called \texttt{frbgeom}. We release it together with posterior samples alongside this paper\footnote{\url{https://dx.doi.org/10.5281/zenodo.16421753}}.

\subsection{Bayesian inference}
\begin{figure}
	\centering
	\includegraphics[width=1\linewidth]{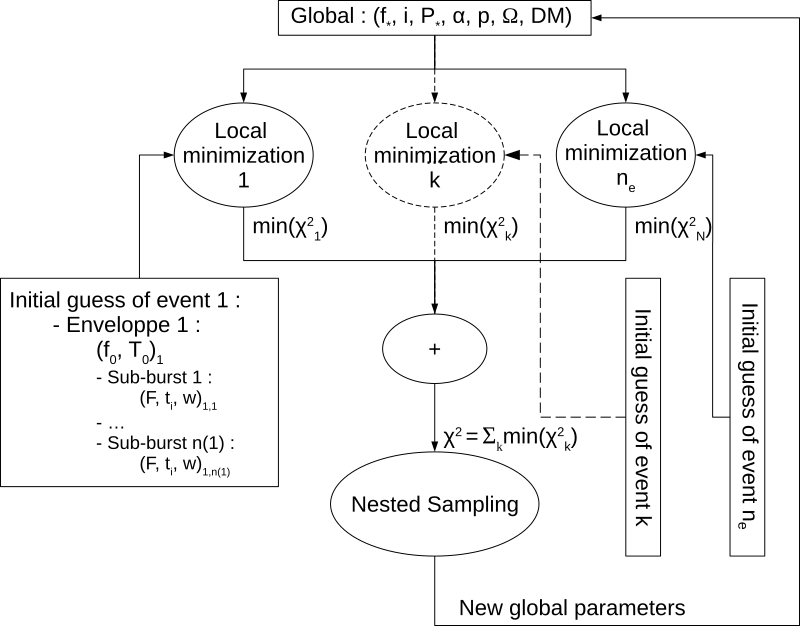}
	\caption{Diagram of the two-stage fitting process. Global parameters are iterated using nested sampling. For a fixed set of global parameters, the $n_e$ envelopes (one per event) each have a set of local parameters comprising two envelope parameters, $f_0, T_0$ (see main text), and parameters for each of the $n(k)$ sub-bursts (also denoted as $n_{b/e}$ in the main text), with $k$ the envelope index. Local parameters were fitted independently to the dynamic spectrum of each event using least-square minimisation. Initial guesses were obtained thanks to pre-processing. }
	\label{fig:diagramfitting}
\end{figure}

We sampled the posterior distribution functions of global parameters using \texttt{Multinest} \citep{feroz_multinest_2009} through the \texttt{PyMultinest} python interface \citet{buchner_x-ray_2014}\footnote{\url{https://github.com/JohannesBuchner/PyMultiNest}}. The main advantage of the nested sampling method is the fact that it explores a bounded region of the parameter space without being a priori biased by a local seed initialisation, as is the case in a typical Metropolis-Hastings Markov chain Monte Carlo. Related to this is its capacity to capture efficiently local modes without remaining trapped in one, which proved particularly helpful in this work.

Due to the large number of parameters involved, 84 for A and 128 for B, it is not possible to explore the full parameter space with reasonable computing resources. However, we used the fact that local parameters are mostly uncorrelated with global ones. Indeed, local parameters only influence their respective events, while global parameters affect the whole dataset (see Sec. \ref{sec:model}). Thus, the data were fitted following two nested stages Fig. \ref{fig:diagramfitting}. First, given a set of global parameters, local parameters were fitted independently on each event's data using a least-square minimiser\footnote{We used the non-linear least-square minimiser of the \texttt{Scipy} python library.}. The global log-likelihood was then obtained as $-\chi^2/2 = -\sum_i \min(\chi^2_i)/2$, where $\min(\chi^2_i)$ are the locally minimised chi-square values of each event. If this can be seen as a form of marginalisation over local parameters, we caution that it is not strictly speaking the case.

Overall, each of the two runs took $\sim 20,000$ CPU hours, running on 96 Intel Xeon Gold 6126 cores. Since the code is currently written in \texttt{python}, there is a potential for a significant speed-up by converting at least some key parts of the code into a compiled language.

The local optimisation, by abruptly reducing the number of dimensions from $\sim 100$ down to $k = 7$, tends to create a more disconnected $\chi^2$ landscape, which is therefore harder to sample. Apart from optimisation itself, the main reason for this disconnection is the systematic noise created by the local fitting procedure. Indeed, there is a significant variance of the optimal local $\chi^2_i$ depending on the initial conditions of the fit. By significant, we mean much larger than the expected standard deviation of the $\chi^2$ distribution for a linear model with $k$ degrees of freedom, which is $\sigma_k = \sqrt{2k}\simeq 3.7$. Any variation $\gg \sigma_k$ is exponentially penalised in the posterior estimations, which can create huge difficulties in the sampling process\footnote{This is because posterior probabilities result from the (Monte Carlo) integration over a $k$-dimensional space and, by definition, these probabilities are non-negligible only for $\chi^2$ values falling within at most a few standard-deviation $\sqrt{2k}$ of the mean value.}. These systematics must be accounted for, both to ensure convergence and to obtain estimates of posterior uncertainties that are sufficiently conservative.

Typically we observe a characteristic dispersion due to the fitting noise, $\sfn < 0.01\chi^2$, considering that $\chi^2 \sim \ndof \sim 10^5$ with $\ndof$ the number of likelihood degrees of freedom. This is equivalent to an identical proportion of bins changing their value by one standard deviation of the background noise. Thus, this is not something that can usually be easily noticeable by eye, and the fit remains qualitatively similar. A more quantitative way of estimating the impact of this systematic noise is to compare it to the standard deviation of the likelihood's $\chi^2$, $\sigma_n = \sqrt{2n} \sim 500$, where $n\simeq\ndof$ is the number of data bins. Thus we see that $\sigma_n\sim\sfn$, meaning that the impact of the fitting noise has a similar amplitude to that of having a different realisation of the background noise. 

\begin{figure}
	\centering
	\includegraphics[width=0.49\linewidth]{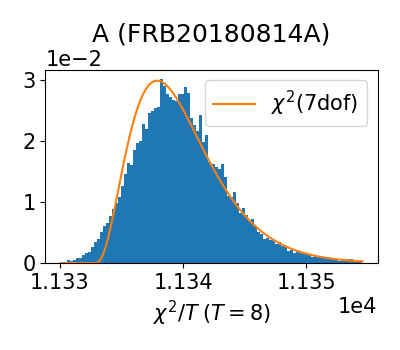}
	\includegraphics[width=0.49\linewidth]{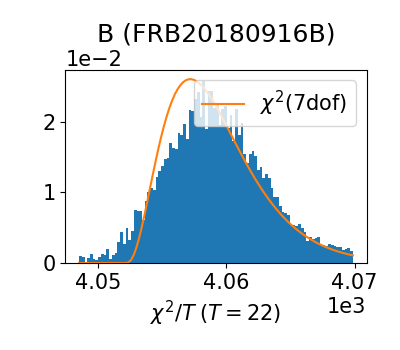}
	\caption{Posterior distribution of $\chi^2$ for source A (left) and B (right). As a reference, the orange lines represent the theoretical $\chi^2$ distribution for a Gaussian posterior with 7 degrees of freedom.}
	\label{fig:chi2distro}
\end{figure}

The solution we adopted to smooth the landscape was to include a temperature, $T$, such that the log-likelihood is $-\chi^2/2T$. The temperature was chosen such that $\mathrm{var}(\chi^2/T) \simeq 2k$, which is the variance of a Gaussian mode with $k$ degrees of freedom. We find this value to smooth the landscape enough for the sampling to perform well. We chose $T = 8$ for A and $T=22$ for B. Figure \ref{fig:chi2distro}  shows that with these temperatures we recover posterior $\chi^2$ samples broadly similar to the $\chi^2$ distribution that would be obtained if the posterior were Gaussian (with seven degrees of freedom). Thanks to this, we expect the sampling to be reasonably well behaved.
Due to large uncertainties on the surface frequency parameter, $f_*$, and the toroidal-to-poloidal ratio, $\alpha$, sampling was performed against $\log_{10} (f_*/\mathrm{MHz})$ and $\log_{10} (\alpha)$, and thus effectively used log priors.

\subsection{Priors} \label{sec:priors}
 Nested sampling requires a finite parameter space to sample. All our priors are flat, once the logarithmic transformation mentioned above is considered; they are summarised in Table \ref{tab:prior}.

The lower bound on $f_*$ simply coincides with the highest observed frequencies in the CHIME/FRB data, which is 800 MHz. The upper bound at $10^4 \rm GHz$ is much higher than any FRB observation, and not reached in any of our runs. 

The lower bound on the spin period is set to approximately three times the duration of the longest event. This is dictated by the locality hypothesis, whereby an emission region can only occupy a fraction of the magnetosphere. The upper bound of $15 ~\rm s$ is somewhat larger than most known confirmed magnetars (see footnote \ref{fn:mcgillcat}), albeit faster than the slowest-spinning neutron stars known \citep{agar_broad-band_2021,caleb_discovery_2022}. Boundaries were not reached in any run.

The prior on $\alpha$ is only positive such that the orientation of the toroidal field can only be counter-rotating. Since we sample against $\log_{10} (\alpha)$, it is impossible to account for both signs simultaneously. However, test runs with linear scaling showed that negative values were thoroughly dismissed. This is readily understood considering that a negative value cannot possibly account for the frequency downward-drifting patterns, as is demonstrated in \citet{voisin_geometrical_2023}. Therefore we decided to set a lower bound at $\alpha > 0.1$ ($0$ being excluded by the logarithmic scaling). The upper bound of $10^4$ far exceeds the magnitude of any inferred none-dipolar magnetic component in magnetars, and was never reached in our runs. 

The lower bound on the toroidal power-law index, $p$, includes the monopole limit, $p=1$, of self-similar force-free solutions \citet{thompson_electrodynamics_2002}. On the other hand, the upper bound of $8$ implies an extremely steep decline of the toroidal field with respect to the dipolar one that would confine it very close to the neutron star unless extreme ratios $\alpha$ were also given. These boundaries were not reached in any run. 

The characteristic opening angle, $\Omega$, must be small by model assumption. An upper bound of $0.5 ~ \rm rad$ fulfils this criterion without being too constraining. It was not reached in any run. 

The DM measurements for individual events in the CHIME/FRB catalogue range between $[188.5, 192.5] \; \rm pc\cdot cm^{-3} $ for A, and $[348.7, 350.2] \; \rm pc\cdot cm^{-3}$ for B. We therefore chose ranges wider than these by a couple of DM units, which appears to be sufficient as these boundaries were not reached in either run.

\subsection{Mode selection}
\begin{figure}
	\centering
	\begin{subfigure}{0.99\linewidth}
		\includegraphics[width=1\linewidth]{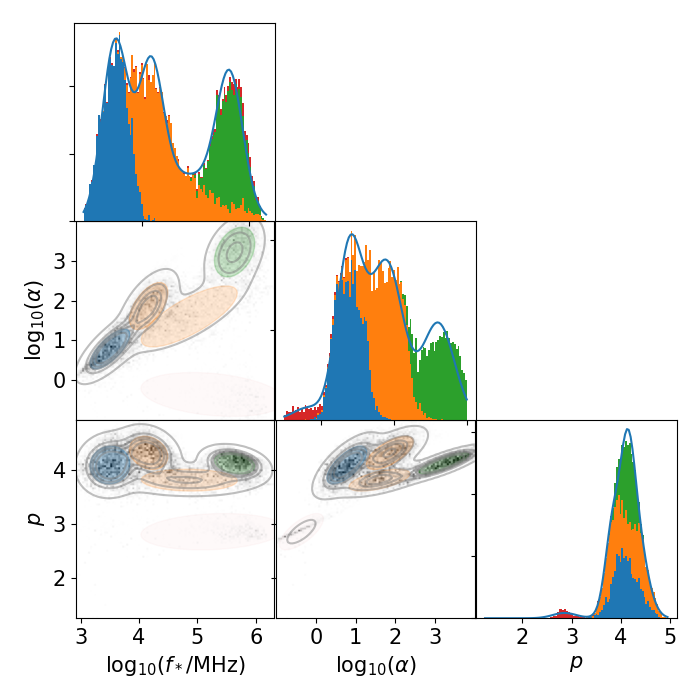}
		\caption{A (FRB20180814A)\label{fig:modesuma}}
	\end{subfigure}
	\hfill
	\begin{subfigure}{0.99\linewidth}
		\includegraphics[width=1\linewidth]{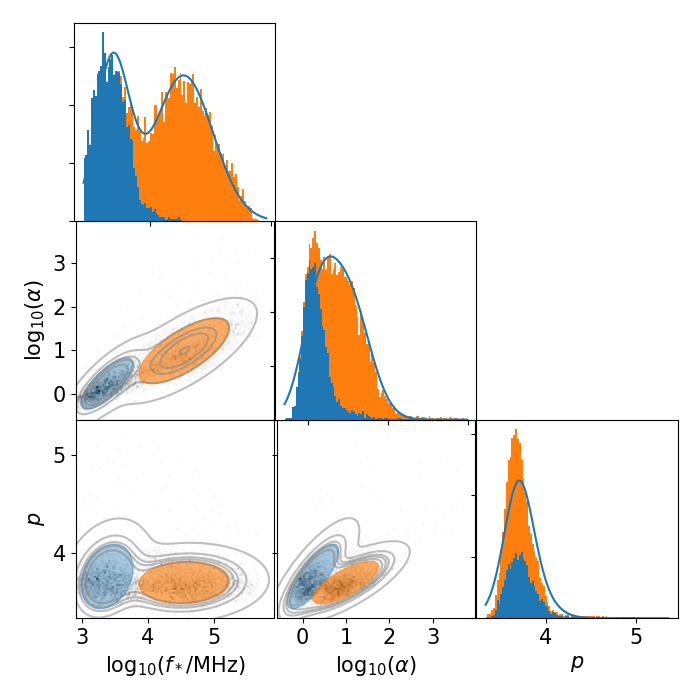}
		\caption{B (FRB20180916B)\label{fig:modesumb}}
	\end{subfigure}
	\caption{Correlation plot of the posterior sample of source A (a) and B (b)  marginalised over all but three parameters:  the surface emission frequency, $\log_{10}(f_*)$, the surface toroidal-to-poloidal magnetic-field ratio, $\log_{10}(\alpha)$, and the radial power-law exponent of the toroidal field, $p$. A Gaussian mixture model has been fitted to this marginalised distribution. On the 2D correlation plots, grey iso-contours of the model are shown together with coloured ellipses representing the one-standard-deviation area of each mode. The same colours are used on the 1D plots to represent the parts of the histogram attributed to each mode, and the line represents the model. In particular, the low mode is blue, the medium mode is orange, and the high mode is green.}
	\label{fig:modesum}
\end{figure}

We used a Gaussian mixture model in order to characterise the modes that appear in the parameter sample. In a marginalised distribution, modes are mostly visible for the surface frequency parameter, $f_*$. However since this one correlates with the magnetic parameters $\alpha$ and $p$, including these generates much better results. At the same time, including the other parameters of the model creates difficulties due to the increasing dimensionality of the problem without significant improvement. The Gaussian mixture was fitted using the expectation-maximisation algorithm\footnote{We used the implementation in \texttt{scikit-learn}, \url{https://scikit-learn.org/}}, the result of which is shown in Fig. \ref{fig:modesum}. The number of components was kept limited such that the weight of an extra component would be less than 2\%. This led to five and two components for A and B, respectively.  

The three main modes of A together represent 97\% of the distribution (see appendix \ref{apsec:modes}), and are labelled as low, medium, and high modes according to the surface frequency, $f_*$, they correspond to (see Fig. \ref{fig:modesuma}). For B, only the low and medium mode are present, representing 100\% of the sample (see Fig. \ref{fig:modesumb}). 
	Each element of the sample can be assigned to a mode based on the probability derived from the Gaussian mixture model. In the case of A, the medium mode (in orange in Fig. \ref{fig:modesuma}) is actually the union of two Gaussian components that share the same qualitative properties, such that discussing them separately would be redundant. These two sub-components are visible as the two orange ellipses in the 2D histograms in Fig. \ref{fig:modesuma}.

\subsection{Maximum transverse size of emission regions} \label{sec:transverse}
For a given observer direction, $\vec{u}$, and a unit vector indicating the direction of the magnetic field, $ \vec{n} = \vec{B}/B$, the characteristic region that can be seen by the observer is characterised by $ \vec{u} \cdot \vec{n} = \cos \Omega$, where $\Omega$ is the characteristic opening of the emission beam. The visible transverse region is the cross-section of the visible region through the plane orthogonal to the magnetic field at the emission site. Assuming that $\Omega$ is fully due to the divergence of field lines, that is that emission is infinitely beamed around the magnetic field direction, then one can define the maximum transverse region as above. 

\begin{figure}
	\centering
	\includegraphics[width=1\linewidth]{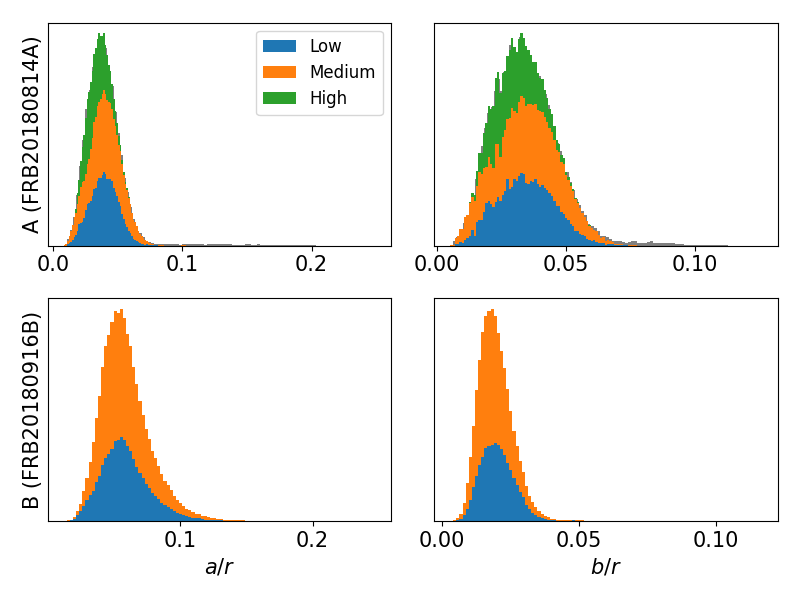}
	\caption{Posterior distributions of transverse sizes: semi-major axis, $a$, and semi-minor axis, $b$, of the approximating ellipses, relative to the radius, $r$. The first row represents values for A, and the second row values for B. For better visibility the left-hand histograms are limited to values of $a/r < 0.25$, which excludes from the plot about $0.1\%$ of all samples for A and $ 0.004\%$ for B. The colours represent modes according the convention of Fig. \ref{fig:modesum}.}
	\label{fig:transverse}
\end{figure}

Expanding $\vec{n}$ to quadratic order in the orthogonal plane, the above equation, $ \vec{u} \cdot \vec{n} = \cos \Omega$, becomes that of an ellipse of the semi-major axis, $a$, and semi-minor axis, $b$. As long as the ellipse is small compared to the typical scale of variation in the magnetic field, which scales as the distance to the centre of the star, the second-order approximation is reasonable. As is shown in Fig. \ref{fig:transverse}, this is generally the case in our study except for a few outliers.

\begin{table}[h]
	\caption{Prior ranges used in MCMC runs of each source. } \label{tab:prior}
	\begin{tabular}{lll}
		& A & B \\
		\hline
		$\mathrm{log_{10}}(f_*/\mathrm{MHz})$   &  $\mathrm{log_{10}}([800, 10^{7}])$ & "\\
		$P_* \rm (s)$   & $[0.180, 15]$  & $[0.280, 15]$ \\ 
		$\mathrm{log_{10}}(\alpha)$   &  $\mathrm{log_{10}}([0.1,10^4])$ & " \\ 
		$p$   & $[0,8]$ & "\\ 
		$\Omega \rm (rad)$   & $[0,0.5]$ & "\\ 
		$\rm DM (cm^{-3}\, pc)$  & $[187,194]$ &  $[347,352]$\\
	\end{tabular}
\tablefoot{Quotes indicate identical priors.}
\end{table}

\section{Results}\label{sec:results}

\begin{table}[h]
	\caption{MCMC result summary for both sources A and B\label{tab:res}}
	\begin{tabular}{lcc} 
		 & A (FRB2018081A) & B (FRB20180916B) \\
		\hline
		$\mathrm{log_{10}}(f_*/\mathrm{MHz})$   &   $4.5_{-1.0}^{+1}$   &   $4.(1)_{-8}^{+7}$\\ 
		$i \rm (rad)$   &   $1.(6)_{-2}^{+2}$   &   $1.8_{-1}^{+0.8}$\\ 
		$P_* \rm (s)$   &   $2.(3)_{-5}^{+5}$   &   $0.8_{-0.2}^{+0.1}$\\ 
		$\mathrm{log_{10}}(\alpha)$   &   $2_{-1}^{+1}$   &   $0.7_{-0.6}^{+0.6}$\\ 
		$p$   &   $4.(1)_{-2}^{+2}$   &   $3.(7)_{-1}^{+1}$\\ 
		$\Omega \rm (rad)$   &   $5_{-1}^{+1} \times 10^{-2}$   &   $5_{-1}^{+1} \times 10^{-2}$\\ 
		$\rm DM (cm^{-3}\, pc)$   &   $1.885(7)_{-5}^{+5} \times 10^{2}$   &   $3.489(0)_{-5}^{+5} \times 10^{2}$\\ 
		\hline
		$\chi^2/N_\mathrm{dof}$   &   $90641 / 84181$   &   $89067 / 114044$\\ 
		\hline 
		$r/R_*$   &   $3_{-3}^{+5} \times 10^{2}$   &   $73_{-69}^{+63}$\\ 
		$a/r$ \space \space | \space \space $a/R_*$   &   $4_{-1}^{+1} \times 10^{-2}$ \space \space | \space \space $13_{-13}^{+16}$   &   $6_{-2}^{+2} \times 10^{-2}$ \space \space | \space \space $4_{-4}^{+4}$  \\ 
		$b/r$  \space \space | \space \space $b/R_*$ &   $3_{-1}^{+1} \times 10^{-2}$ \space \space | \space \space $10_{-10}^{+14}$ &   $2.(0)_{-5}^{+5} \times 10^{-2}$  \space \space | \space \space  $1_{-1}^{+1}$ 
	\end{tabular}
\tablefoot{First block, median parameters with a median 68\% confidence interval for both sources A and B. 
	Second block, best $\chi^2$ relative to the number of degrees of freedom.
	Third block, derived parameters including emission height, $r$, and transverse size of the emission region relative to the distance, $r$ (semi-major and semi-minor axis  of the approximating ellipse, $a$ and $b$, respectively). Error bars apply do digits between parentheses.}
\end{table}

\begin{table}
	\caption{Same as Table \ref{tab:res} but limited to the medium mode.} \label{tab:commonmode}
	\begin{tabular}{lcc}
		& A (FRB2018081A) & B (FRB20180916B) \\
		\hline
		$\mathrm{log_{10}}(f_*/\mathrm{MHz})$   &   $4.(5)_{-4}^{+5}$   &   $4.(6)_{-4}^{+4}$\\ 
		$i \rm (rad)$   &   $1.(6)_{-2}^{+3}$   &   $1.9_{-1}^{+0.7}$\\ 
		$P_* \rm (s)$   &   $2.(2)_{-5}^{+5}$   &   $0.8_{-0.1}^{+0.1}$\\ 
		$\mathrm{log_{10}}(\alpha)$   &   $1.(8)_{-4}^{+4}$   &   $1.(0)_{-4}^{+4}$\\ 
		$p$   &   $4.(1)_{-3}^{+3}$   &   $3.(7)_{-1}^{+1}$\\ 
		$\Omega \rm (rad)$   &   $5_{-1}^{+1} \times 10^{-2}$   &   $5_{-1}^{+1} \times 10^{-2}$\\ 
		$\rm DM (cm^{-3}\, pc)$   &   $1.885(7)_{-5}^{+5} \times 10^{2}$   &   $3.489(0)_{-5}^{+5} \times 10^{2}$\\ 
		\hline
		$r/R_*$   &   $1.7_{-1.5}^{+0.3} \times 10^{2}$   &   $1.(2)_{-9}^{+9} \times 10^{2}$\\ 
		$a/r$ \space \space | \space \space $a/R_*$   &   $4_{-1}^{+1} \times 10^{-2}$ \space \space | \space \space  $5_{-4}^{+2}$ &    $6_{-2}^{+2} \times 10^{-2}$ \space \space | \space \space $7_{-5}^{+5}$  \\ 
		$b/r$ \space \space | \space \space $b/R_*$   &   $3_{-1}^{+1} \times 10^{-2}$  \space \space | \space \space $4_{-3}^{+2}$  &  $1.(9)_{-5}^{+5} \times 10^{-2}$  \space \space | \space \space  $2_{-2}^{+2}$  \\ 	
	\end{tabular}
\end{table}

\subsection{General parameters} 
Table \ref{tab:res} summarises the results of our MCMC sampling (see Sec. \ref{sec:methods}), and in particular the seven global parameters described in Sec. \ref{sec:model}. Correlation plots of posterior probabilities are available in appendix \ref{apsec:corner} and best-fit residuals are available on the \href{https://dx.doi.org/10.5281/zenodo.16421753}{Zenodo repository}. Interestingly, both sources give broadly similar results. 

An example of a best-fit residual for a particular event is given in Fig. \ref{fig:residuals} (see the \href{https://dx.doi.org/10.5281/zenodo.16421753}{Zenodo repository} for the others). We caution that the fit was performed on the set of all events, and therefore it cannot be judged by looking at a singe event. In the particular case of Fig. \ref{fig:residuals}, one can see that the four components are following a clear downward-drifting pattern that is well within the one-$\Omega$ envelope. This is not always the case, as a burst can be seen from the outer part of the beam (see Sec. \ref{sec:model}, and the \href{https://dx.doi.org/10.5281/zenodo.16421753}{Zenodo repository} for examples). Moreover, if the fit appears here to be visually quite satisfactory, there are two events (out of 19) from B for which this is clearly not the case. This suggests that a more complex geometry is involved. One of them is FRB20181019A, which is the only one where the two components are fully separated by a relatively long 60 ms emission gap (see also Sec. \ref{sec:data}). One may think that two emission regions are at play, in which case this event could be split in two with two different envelopes. The other ill-fitting event is FRB20190605B.

In both cases, the source DM is well determined and lies within the interval of per-event DMs calculated in the CHIME/FRB catalogue. 
Inclinations, $i$, are distributed nearly symmetrically around $90^\circ$ (see appendix \ref{apsec:corner}). The opening angle, $\Omega$, is well determined with values of $5_{-1}^{+1}\times 10^{-2}  ~ \rm rad$ ( $\sim3^\circ$) in both cases, which is compatible with the model's assumption that $\Omega \ll 1 ~ \rm rad$.

The spin periods are the main difference between A and B, with $P_* = 2.3_{-0.5}^{+0.5} ~ \rm s$ and $P_* = 0.8_{-0.1}^{+0.2} ~ \rm s$ respectively. This parameter moderately correlates with $\Omega$ and $p$ (see appendix \ref{apsec:corner}), but the shorter period of B is caused by the stronger rate of downward drifting observed from this source \citep{voisin_geometrical_2023}.

The magnetic parameters, $\alpha,p $, and surface emission frequency, $f_*$, are correlated (Fig. \ref{fig:modesum}). In particular, the toroidal-to-poloidal ratio, $\alpha$, visibly correlates with $p$ and $f_*$. Its values are $\log_{10} \alpha = 2_{-1}^{+1}$ and $\log_{10} \alpha = 0.7_{-0.6}^{+0.6}$ for A and B, respectively, compatible with the two sources thanks to the relatively wide uncertainties. The toroidal index, $p$, is well determined and similar for the two sources, with values of $p = 4.(1)_{-2}^{+2}$  and  $p = 3.(7)_{-1}^{+1}$ for A and B, respectively. The surface emission frequency, $f_*$, is in the range of $10 - 40 ~ \rm GHz$ for A and $15 - 100 ~ \rm GHz$ for B. 

\subsection{Emission sites} 
\begin{figure}
	\centering
	\includegraphics[width=1\linewidth]{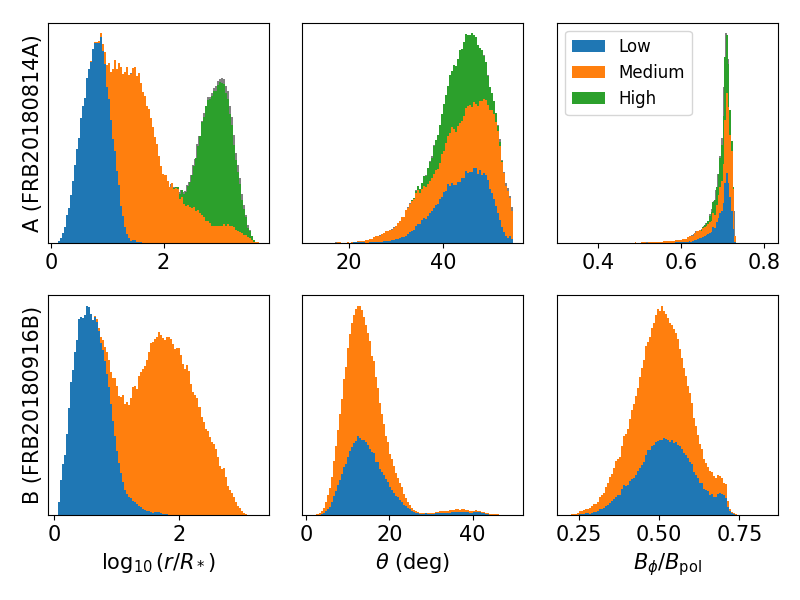}
	\caption{Posterior distributions of radius $r$, colatitude $\theta$, and toroidal-to-poloidal magnetic ratio of emission site for source A, top row, and source B, bottom row. The colours represent modes according the convention of Fig. \ref{fig:modesum}.}
	\label{fig:pos}
\end{figure}

Figure \ref{fig:pos} shows the distribution of emission sites in the magnetosphere averaged over all events (each event has its own posterior distribution). The altitude is broadly distributed between a few stellar radii up to about 1000, which, given the spin periods involved, is compatible with the model assumption of a distance, $r$, sufficiently low to assume co-rotation. Colatitudes are relatively well localised, respectively, around $45^{\circ}$ and $15^{\circ}$ for A and B. The ratio between the toroidal and poloidal components of the magnetic field at the emission site is narrowly distributed between 0.3 and 0.8. Since this ratio is directly related to the narrow-band downward-drifting pattern, we indeed expect a relatively stable value across the parameter sample. It is notably in line with \citet{voisin_geometrical_2023}, which proposed that a value of order one was necessary to reproduce the observed features of repeaters. 

The characteristic transverse sizes of the emission regions (see Sec. \ref{sec:transverse}) appear to be small, with ratios of the semi-major axis to the radius, $a/r$, and semi-minor axis to the radius, $b/r$, that are typically of the order of $\sim \text{a few} \times 10^{-2}$ ( Table \ref{tab:res} and Fig. \ref{fig:transverse}). As before, we considered size distributions averaged over all events. Such a limited extent is compatible with the model assumption of locality. The semi-major axis $a$ typically is not larger than a few times the semi-minor axis $b$, indicating that the emission ellipse can be quite flat. Mean semi-major axis are between $13_{-13}^{+16} R_*$ for A and $ 4_{-4}^{+4} R_*$ for B (Table \ref{tab:res}), with the 99th percentile of the distribution reaching $1.4 \times 10^{2} R_*$ for A and $33 R_*$ for B.

\subsection{Modes} 
\begin{figure}
	\centering
	\includegraphics[width=0.49\linewidth]{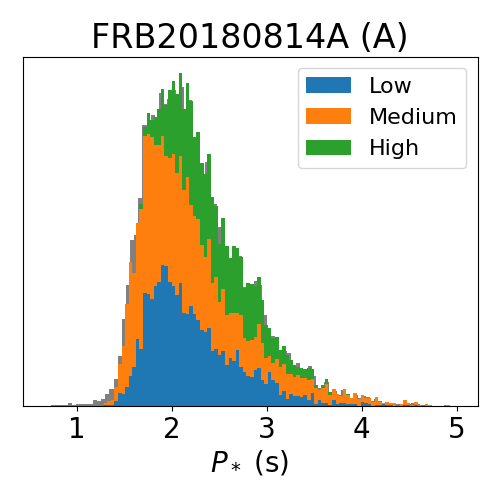}
	\includegraphics[width=0.49\linewidth]{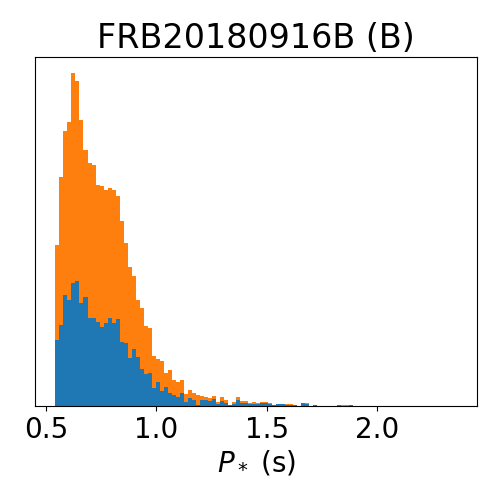}
	\caption{Posterior distributions of the spin periods for FRB20180814A (A), left-hand side, and FRB20180916B (B), right-hand side. The colours represent modes according to the convention of Fig. \ref{fig:modesum}.}
	\label{fig:Pspin}
\end{figure}

Posterior distribution functions can be decomposed into three main modes corresponding to low, medium, and high surface emission frequencies, $f_*$, corresponding, respectively, to  $f_* \sim 3.5 ~ \rm GHz$, $\sim 10 - 100 ~ \rm GHz$, and $\sim 500 ~ \rm GHz$ (see Fig. \ref{fig:modesum}). Source A shows all three modes, detailed in the extended data in Table \ref{tab:frbAmodpars}, while B shows only the low and medium ones (Table \ref{tab:frbBmodpars}). These modes  also differ by their magnetic surface ratios, $\alpha$. Orders of magnitudes are $\alpha \lesssim 10$, $10 \lesssim \alpha \lesssim 100$, and $100 \lesssim \alpha \lesssim 10^4$ for the low, medium, and high modes, respectively. Somewhat larger values for A are found relative to B. The spin period, in particular, is similar for each mode (see Fig. \ref{fig:Pspin}).

\section{Discussion}\label{sec:disc}

\subsection{Preferred mode}
\label{sec:discussionmode}
The presence of multiple modes is at least partly a consequence of the high degree of symmetry of the model. This allows distinct regions of the magnetosphere, and in particular different altitudes, to match the line of sight. We thus have three main modes corresponding to low, medium, and high altitudes (see Fig. \ref{fig:pos}).

Prior knowledge can be mobilised to select a preferred mode. So far one source, FRB 20121102, has been observed almost up to $8 ~ \rm GHz$ \citep{gajjar_highest_2018}. If this result could be generalised to A and B, then it would discard the lower mode by setting a lower limit on $f_*$. A second argument for discarding this mode lies in the correspondingly low altitudes of the emission regions, less than $\sim 10 R_*$, which reduces the chances of radiation escaping the magnetosphere \citep{beloborodov_can_2021,huang_fast_2024, qu_transparency_2022}. 

The medium mode is compatible with observational limits in surface frequency with an overall interval comprised between $10 \,\mathrm{GHz}$  and $100 \,\mathrm{GHz}$, and altitudes broadly distributed around $\sim 100 R_*$ (Fig. \ref{fig:pos}, appendix \ref{apsec:modes}). Interestingly, recent numerical work shows that FRBs might be sourced by the dissipation of shocks in the same altitude range, assuming typical magnetar settings \citep{beloborodov_monster_2023, vanthieghem_fast_2025}. The medium mode also has the largest weight in the Gaussian mixture of both sources (see Tables \ref{tab:frbAmodpars} and \ref{tab:frbBmodpars}).

The high mode of source A corresponds to emission altitudes of $\sim 1000 R_*$ (Fig. \ref{fig:pos}, Table \ref{tab:frbAmodpars}), and predicts an upper frequency much higher than currently observed unless a cut-off at a fairly high altitude is additionally assumed. In addition, this mode corresponds to a surface magnetic-field ratio of $\log_{10}(\alpha) = 3.2_{-0.4}^{+0.4}$ $ (6\times 10^2 < \alpha < 4.0 \times 10^3$, appendix \ref{apsec:modes}), the central value of which is almost an order of magnitude larger than the largest inferred ratios between non-dipolar and dipolar components from observations of two low-field magnetars \citep{tiengo_variable_2013, rodriguez_castillo_outburst_2016}, while the medium and low mode are both similar to or under this value. Additionally the fact that the medium mode is common to both sources is somewhat expected if one assumes that their similar behaviour arises from a similar physical configuration. Indeed all parameters are compatible within uncertainties except for $P_*$ and DM, meaning that both magnetic configurations are similar. For all these reasons, we suggest that the medium mode is more likely than the two others given prior knowledge, and report its detailed parameters in Table \ref{tab:commonmode}. 
Nonetheless, in the remainder of this section we conservatively quote the general values of the parameters (Table \ref{tab:res}), unless otherwise stated. This does not affect the conclusions, and the reader may  compare these values with the more accurate ones in Table \ref{tab:commonmode} that are associated with the common mode.

\subsection{Magnetic configuration} \label{sec:magconf}
It is quite remarkable that the toroidal magnetic index is $p \sim 4$ for both sources. This value is evocative of the quadrupolar order of a multi-polar decomposition of the magnetic field in vacuum. Vacuum configurations are relevant because the magnetic field is expected to strongly dominate the plasma at low altitudes in most of the magnetosphere. In addition, the quadrupole order is more likely to be detected at intermediate altitudes than other multipoles due to its slower decay with distance. However, vacuum quadrupolar components cannot create a homogeneous toroidal field as assumed here, but could contribute to it under suitable lines of sight.

In known semi-analytical self-similar solutions of twisted axisymmetric magnetospheres \citep{thompson_electrodynamics_2002, wolfson_shear-induced_1995}, the toroidal and poloidal components share the same radial dependence, whereby the toroidal component results from magnetospheric currents. In these particular solutions, integer exponents, $p$, correspond to vacuum solutions, and are thus without a toroidal component. Non-integer values correspond to twisted magnetospheres \citep{wolfson_shear-induced_1995}. A recent solution \citep{voisin_twisted_2025} does possess the satisfying combination of an aligned dipole with a toroidal component decaying as $1/r^4$. However, the surface toroidal-to-dipolar ratio is about 1 in these solutions. 

Recent numerical approaches \citep{vigano_force-free_2011, fujisawa_magnetic_2014, pili_general_2015, glampedakis_inside-out_2014, akgun_long-term_2017, uryu_equilibriums_2023} simultaneously solve for the stellar interior and the magnetosphere of extremely magnetised stars, and show a mix of toroidal and poloidal components concentrated near the stellar surface in the equatorial region, as well as a dipole-like poloidal field on larger scales. Indeed, except in \citet{uryu_equilibriums_2023}, the twist is confined within a given magnetic field line.
In all these cases, the ratio between poloidal and toroidal components does not exceed $\sim 3$ at the stellar surface (Uryu, private communication 2024). Moreover, the surface ratio remains somewhat smaller than the lower end of the range found in this paper ($10<\alpha<1000$ for A and $1.3<\alpha<20$ for B at a 68\% confidence level, but, respectively, $25<\alpha<158$ and $4<\alpha<25$ for the common mode) . More generally, the limited surface toroidal-to-dipolar ratio is intrinsically related to the existence of a maximum twist in force-free solutions \citep[e.g.][]{akgun_force-free_2016, mahlmann_instability_2019}. 

Thus, if the results of our phenomenological model are evocative of what may be called a twisted quadrupole, the discrepancy with available self-consistent magnetospheric solutions is significant and work is needed to reconcile both approaches. Notably, fitting multi-polar magnetic configurations or going beyond force-free solutions might be necessary.

More generally the large ratio between dipolar and non-dipolar magnetic components suggests a relatively low amplitude of the dipolar component. Indeed, assuming a maximum surface field of $\sim 10^{11} \rm \;\mathrm{T}$ ($10^{15}$G), the dipolar component should then be $B_{\rm dip} \lesssim 10^{10} \rm \;\mathrm{T}$ ($10^{14}$G). This is evocative of low-field magnetars, which are characterised by a relatively small dipolar component but large multi-polar ones. In particular, the three known low-field magnetars are SGR 0418+5729 with a surface dipolar strength of $B_{\rm dip} \simeq 6\times 10^{12} \rm \;\mathrm{T}$ ($6\times 10^{16}$G) \citep{rea_low-magnetic-field_2010, rea_outburst_2013}, SWIFT J1822.3-1606 with $B_{\rm dip} \simeq 3\times 10^9 \rm \;\mathrm{T}$ ($3\times 10^{13}$G) \citep{scholz_post-outburst_2012, rea_new_2012}, and 3XMM J185246.6+003317 with $B_{\rm dip} \simeq 4.1\times 10^9 \rm \;\mathrm{T}$ ($4.1 \times 10^{13}$G) \citep{rea_3xmm_2014, zhou_discovery_2014}. The first two show spectral lines that, if interpreted as cyclotron resonances, indicate multi-polar components that are one to two orders of magnitude stronger than the dipolar one.

\subsection{Emission regions, beaming, and minimum Lorentz factor}
Interestingly, the emission altitude is consistent with a pulsar-like emission mechanism from low-field magnetars or some pulsars. Indeed, \citet{gil_period_1993} have shown that radio emission starts at altitudes where the magnetic field drops below 1000T ($10^7$G) across all types of pulsars. Thus, assuming that $B < 1000\;\rm \mathrm{T}$  at the emission site at $r\sim 100 R_*$  (Table \ref{tab:res}-\ref{tab:commonmode} and Fig. \ref{fig:pos}), we obtain a surface field of $B_{\rm dip} < 10^9 \;\mathrm{T}$ ($10^{13}$G). The toroidal field has been neglected in this order-of-magnitude estimate as $B_{\phi} < B_{\rm pol}$ at the emission site (Fig. \ref{fig:pos}). Such dipolar strength is compatible with low-field magnetars (see also above) in a way that is noticeably consistent with the above discussion on the toroidal-to-dipolar field ratio, $\alpha$.

The question of whether the emission occurs on open field lines originating from the polar cap, or on closed field lines, is important to the emission physics. The polar-cap angular radius at altitude $r$ is $\theta_{\rm pc} \simeq \sqrt{r/R_{\rm LC}}$ with $R_{\rm LC} = cP_*/2\pi$ the light-cylinder radius\footnote{This is the usual formula for a dipole, but it is here unaffected by the toroidal field due to axial symmetry.}. If the colatitude of the emission region is such that $\theta < \theta_{\rm pc}$, then the emission originates from the open field-line region. In our posterior sample, the probability that it occurs is $\simeq 0.2\%$  for A and $\simeq 15\%$ for B. The rest of the sample is well within the closed-field-line region. For A, emission in the open region only happens for the highest altitudes $\sim 1000 R_*$ while it occurs at nearly all altitudes for B thanks to correlations with spin and colatitudes. Although this would appear to disfavour models where emission occurs on open field lines, we caution that our assumption of an aligned dipole prevents us from coming to any conclusions, unless perhaps we were to assume a priori a model requiring near-alignment \citep{beniamini_role_2025}.

Both sources, interestingly, share the same beam opening angle, $\Omega \simeq 0.05 \rm rad$. Since the proper frame of emission is assumed to be ultra-relativistic with respect to the observer, the beam angle is the combination of the relativistic beaming angle and of the streamline divergence. Neglecting the latter, one can estimate a lower limit on the opening angle such that $\Omega \gtrsim 1/\gamma$, which translates into a minimum plasma Lorentz factor of $\gamma \gtrsim 20$. This is compatible with the limit of $\gamma \gtrsim 930 (\epsilon/0.01)^{-1/2} (r/R_*)^{-1/2} P_*^{1/2} $ obtained in \citet{beniamini_role_2025}, where $\epsilon$ quantifies the polarisation degree, which results from constraints on beam coherence within a polar-cap geometrical model that makes different assumptions from ours.

On the other hand, assuming that the value of $\Omega$ is dominated by streamline divergence, we can estimate the characteristic transverse size of the emission region (see Sec. \ref{sec:methods}); that is, the size corresponding to an opening of one $\Omega$. Due to the Gaussian nature of the angular emission profile, an observer may receive the emission from at most a few $\Omega$. Approximating the  transverse cross-section of the emission to an ellipse, we find semi-major axes that typically do not exceed 10\% of the altitude of the emission site (Table \ref{tab:res} and Fig. \ref{fig:transverse}). This validates the assumption of locality of the emission region. 

In absolute terms, the maximum characteristic transverse size is about $a \lesssim 2\times 10^{3} \; \mathrm{km}$ for A and  $a \lesssim 4\times 10^{2} \; \mathrm{km}$ for B (assuming $R_* = 12\;\rm km$), with a 99\% probability in both cases (see also Table \ref{tab:res}). We conclude that the transverse size of the emission regions is generally smaller than a few thousand kilometres. This is compatible with the results in \citet{nimmo_magnetospheric_2025}, which used scintillation to constrain the transverse extent of the emission region to be smaller than $3\times 10^4$ km.

\subsection{Spin and effect of spin-down}
The spin periods found in this work, $2.3_{-0.5}^{+0.5}\,\rm s$ for A and $0.8_{-0.2}^{+0.1}\, \rm s$ for B (Table \ref{tab:res}, Fig. \ref{fig:Pspin}), place these sources among the fastest known magnetars. 
Indeed, only two neutron stars with magnetar-like behaviours have a spin period below 2s \citep{olausen_mcgill_2014}\footnote{\label{fn:mcgillcat}  McGill Magnetar catalogue:  \url{http://www.physics.mcgill.ca/~pulsar/magnetar/main.html}}, namely PSR J1846-0258 \citep{livingstone_post-outburst_2011} and Swift J1818.0-1607 \citep{esposito_very_2020} at $\sim 0.3\, \rm s$ and $\sim 1.4 \, \rm s$, respectively. A large majority of magnetars spin with periods between 5 and 12s \citep{olausen_mcgill_2014}. PSR J1846-0258 is peculiar as it is a very young neutron star, $\sim 800 \, \rm yr$, which mostly behaves like a pulsar with occasional magnetar-like outbursts, and may be seen as a transition object.  
It is remarkable that, for B, the $\sim 1 \,\rm s$ spin period  matches the one inferred from the empirical scaling law relating sub-pulse quasi-periods to spin periods exhibited in \citet{kramer_quasi-periodic_2023}. 

The spin period decay timescale, $\tau = P_*/ 2\dot P_*$, caused by the electromagnetic braking of the dipolar magnetic field, is given by $\tau \simeq 46\,\mathrm{yr} \, P_*^2 (B_*/3\times 10^{10} \rm T)^{-2}$, where $B_*$ is the surface value of the dipole  \citep[e.g.][]{lyne_pulsar_2012}. This leads to $\tau \simeq 550 \rm yr$ for A and  $\tau \simeq 66\rm yr$ for B, respectively, using median values of Table \ref{tab:res}. Here we used a fiducial value of $B_*$ about the median of known magnetars \citep{olausen_mcgill_2014}. Since burst morphology and in particular the downward-drifting rate depends on the spin period, the model predicts that morphology will evolve on the timescale given by $\tau$. Such effects might therefore be observable in the near future. In addition, burst morphology may also evolve due to magnetic reconfigurations. 

On the other hand, if the progenitors are somewhat low-field magnetars (see Sec. \ref{sec:magconf}), their characteristic ages might be two orders of magnitude larger. In this case, we do not expect the effect of spin-down on burst morphology to be significant in the coming decades. We note that the three known low-field magnetars have much longer characteristic ages than the above estimates, of the order of $10^5$ to $10^6$ years \citep{rea_3xmm_2014,rea_outburst_2013, rodriguez_castillo_outburst_2016}. However, it has been proposed that low-field magnetars can be born as such when the magnetic field is amplified by the Tayler-Spruit dynamo, which is triggered when the star is spun up by fallback material \citep{barrere_new_2022, igoshev_connection_2025}.

\subsection{Radius-to-frequency mapping}
\citet{thorsett_frequency_1991} showed that the broadband evolution of the pulse profile can be fit by a continuous law of $f \propto 1/(r-r_0)^\beta$, where $r_0$ is a constant and $\beta \sim 1$, although with variations across pulsars. Thus choosing $\beta=1$ corresponds to the low-frequency asymptote of this empirical law. However, we caution that this type of radius-to-frequency mapping cannot explain the profile of a non-negligible number of pulsars \citep{posselt_thousand-pulsar-array_2021} and thus is not general.

If one considers that the emission results from curvature radiation then $f\propto 1/\rho$, where $\rho$ is the curvature radius \citep[e.g.][Eq. 18]{ruderman_theory_1975}. In a magnetic field dominated by its toroidal component then $\rho \sim r$ as field lines asymptotically tend to be circular. Of course, this assumes that all particles have the same energy, which might be locally true, but is certainly too naive in general. 

A larger value of $\beta$ would likely lead to emission taking place very close to the stellar surface, at a few stellar radii, because of the steep decrease in the emission frequency with distance. That behaviour was seen during the inconclusive run with $f = f_* (R_*/r)^p$ (see Sec. \ref{sec:model}). In principle, a larger distance would nonetheless be possible if it were associated with extreme values of the surface emission frequency, $f_*$. Either way, low-altitude emission or very large $f_*$ both appear rather unlikely from a physical point of view, as was discussed above in Sec. \ref{sec:discussionmode}.

Most importantly, further work should aim to constrain the spectral part of the model, at least by fitting for the parameter $\beta$ and possibly using different types of laws to connect spectral properties with geometry. This will, however, likely require the use of a broadband dataset. 

\subsection{Caveats}

The model in this paper relies on many assumptions that will need to be tested. The assumption that emission tangents magnetic field lines is uncertain. It appears to be correct at least for those pulsars for which the rotating vector model functions. However, it does not always work in pulsars, and the emission mechanism might be different in FRB sources. In this respect, an analysis of polarisation with our model might bring useful complementary information \citep{voisin_geometrical_2023}.

It is in principle perfectly possible within the model to use a geometry not based on magnetic field lines; however, this might require going beyond geometry and solving the plasma physics. 
If using the magnetic geometry is correct, the particular geometry we have used in this work might nonetheless be vastly oversimplified. The dipole is unlikely to be aligned, the toroidal field is usually concentrated near the equator in known solutions, multi-polar components might be prominent close enough to the star, and plasma effects can distort the field. Implementing more realistic geometries is certainly a way to improve the model.

In this work the spectrum of an emitting element is essentially a dirac delta function at the frequency given by radius-to-frequency mapping. One expects a broader spectrum such that several altitudes would blend at a given frequency. In addition, radius-to-frequency mapping might not apply, although one can hope for it to be a reasonable approximation locally. In this respect, the model could be improved using physically motivated emission and propagation models.

Finally, the locality assumption might fail in some circumstances. This might be the case for example for FRB20191221A, for which a 217 ms periodic signal has been observed for 3 s \citep{andersen_sub-second_2022}. If this were to be associated with a rotating polar-cap emission, then the locality approximation would not apply due to the quickly varying geometry around the pole (or a sequence of local models would be necessary). Another possibility is that emission remains local, but several distinct regions contribute at the same time. This is a rather straightforward evolution of the model.

\section{Conclusion}
In this work, we have fitted the data from the first CHIME/FRB catalogue for the two repeating sources FRB20180814A (A) and FRB20180916B (B) with the geometrical model presented in \citet{voisin_geometrical_2023}. This model associates burst morphologies with the general properties of the source, such as its spin.

For both sources we have inferred a fast spin, of the order of $\sim 1 - 2$ s, and a high toroidal magnetic field, from a few to a hundred times the magnitude of the dipolar field at the surface, decaying as $\sim 1/r^4$. Based on these results, one may speculate that the relatively high degree of repetition of these sources is to be associated with a relatively young star, as can be deduced from its fast spin, and with a particularly twisted magnetosphere that is very actively releasing its energy.

As a by-product, we obtained the location and size of the emission regions, typically located at $\sim 100R_*$ and with characteristic sizes of $\lesssim 1000 \; \rm km$. Such size is compatible with recent observations \citep{nimmo_magnetospheric_2025}. We also obtained a conservative lower limit on the Lorentz factor of the emitting plasma of $\gamma \geq 20$.

Assuming a typical magnetar dipolar field, we obtained a spin-down age of the order of a few centuries and possibly as short as a few decades. In the latter case, a change in burst morphology could be detectable in the near future. However, the large toroidal-to-poloidal ratio as well as the typical emission altitudes are evocative of low-field magnetars, in which case the characteristic spin-down ages could be up to two orders of magnitude larger. This exploratory work can be improved by using a more realistic model of the magnetic configuration, using a larger dataset with broad-band data in order to constrain radius-to-frequency mapping, and including polarisation as is proposed in \citet{voisin_geometrical_2023}.

\section{Data availability}
Codes, \texttt{Multinest} analysis results, and additional residual plots are available on the Zenodo repository at 
\url{https://dx.doi.org/10.5281/zenodo.16421753}.

\begin{acknowledgements}
The authors would like to thank Dr. Fabrice Mottez and Dr. Marilyn Cruces for helpful discussions and reviews. \\
The authors thank the referee for their insightful review. \\
This work was granted access to the HPC resources of MesoPSL financed
by the Region Ile de France and the project Equip@Meso (reference
ANR-10-EQPX-29-01) of the programme Investissements d'Avenir supervised
by the Agence Nationale pour la Recherche. 
\end{acknowledgements}

\bibliographystyle{aa}
\bibliography{aa55207-25corr}
\begin{appendix}

\onecolumn

\section{Correlation plots of posterior probabilities}\label{apsec:corner}

\begin{figure}[h]
	\centering
	\includegraphics[width=1\linewidth]{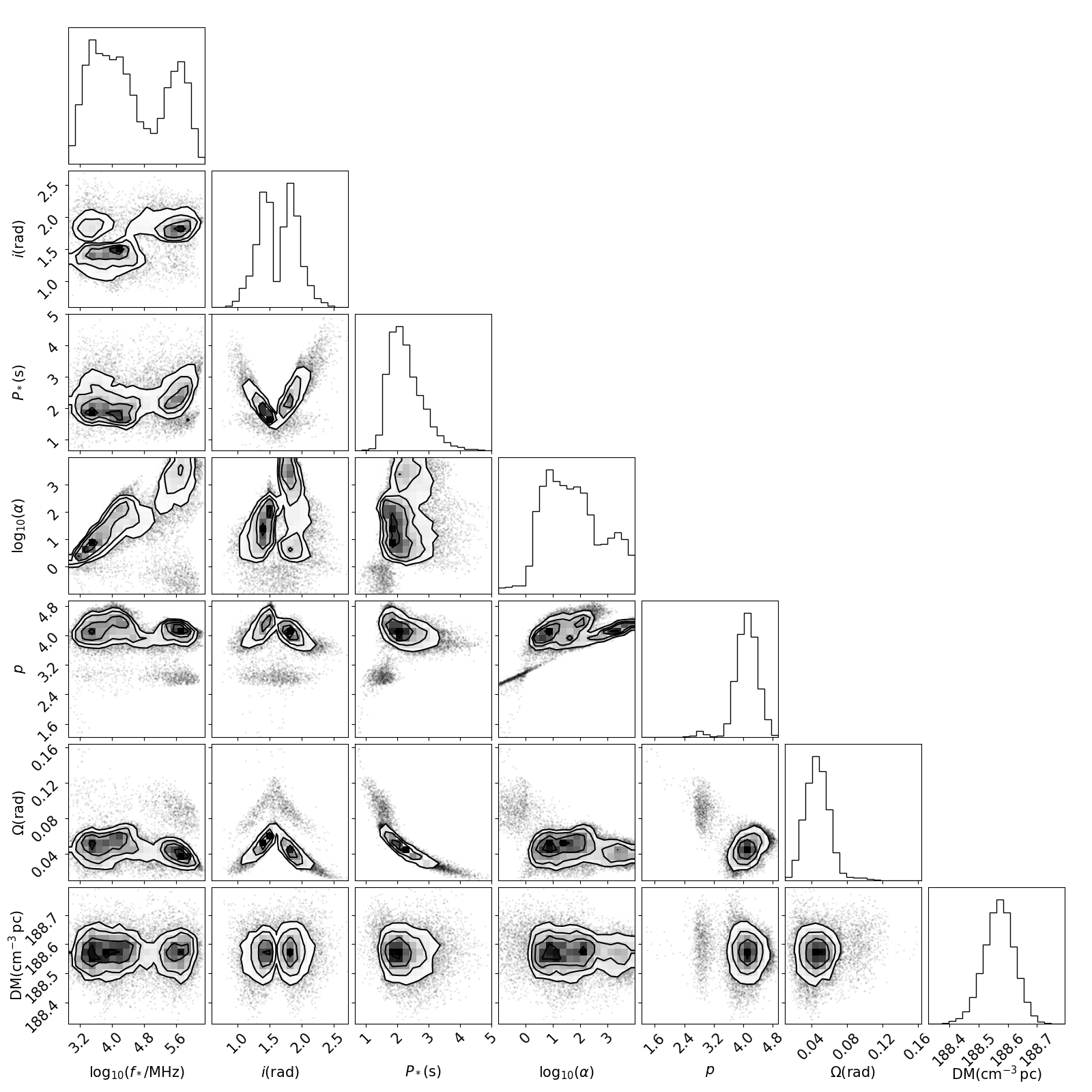}
	\caption{Correlation plot of the posterior distribution of parameters for source A.}
	\label{fig:corner180814beta1}
\end{figure}

\FloatBarrier
\begin{figure*}[!ht]
	\centering
	\includegraphics[width=1\linewidth]{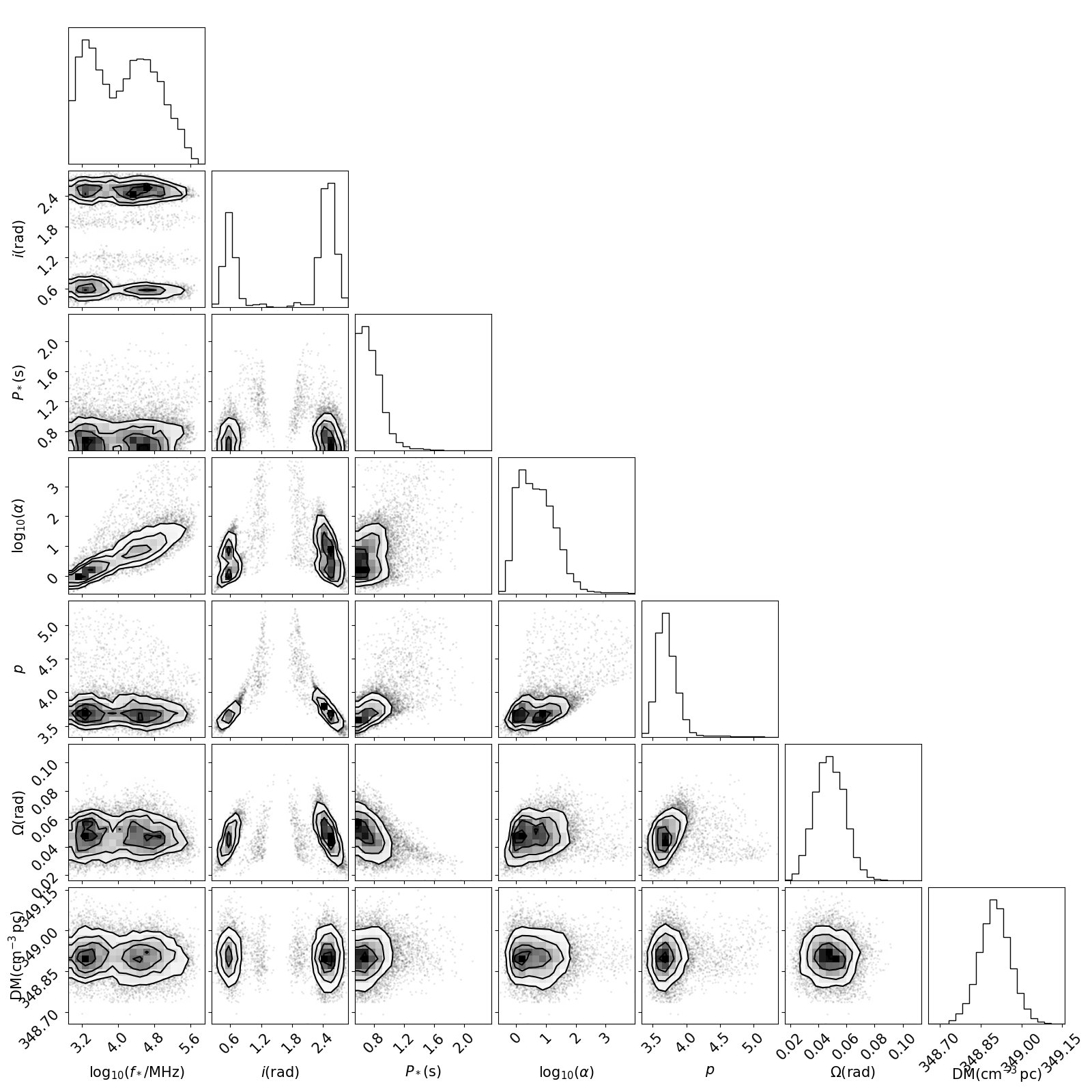}
	\caption{Correlation plot of the posterior distribution of parameters for source B.}
	\label{fig:corner180916beta1}
\end{figure*}

\FloatBarrier
\clearpage

\section{Modes} \label{apsec:modes}

\begin{table*}[h]
	\caption{MCMC result summary for source A, decomposed by mode. } \label{tab:frbAmodpars}
	\begin{tabular*}{\textwidth}{@{\extracolsep\fill}lcccc}

		Weight   &   $0.32$   &   $0.42$   &   $0.24$\\ 
		\hline
		$\mathrm{log_{10}}(f_*/\mathrm{MHz})$   &   $3.(5)_{-2}^{+2}$   &   $4.(5)_{-4}^{+5}$   &   $5.(6)_{-2}^{+2}$\\ 
		$i \rm (rad)$   &   $1.(6)_{-2}^{+3}$   &   $1.(6)_{-2}^{+3}$   &   $1.8(2)_{-9}^{+9}$\\ 
		$P_* \rm (s)$   &   $2.(2)_{-4}^{+4}$   &   $2.(2)_{-5}^{+5}$   &   $2.(5)_{-4}^{+4}$\\ 
		$\mathrm{log_{10}}(\alpha)$   &   $0.8_{-0.4}^{+0.4}$   &   $1.(8)_{-4}^{+4}$   &   $3.(2)_{-4}^{+4}$\\ 
		$p$   &   $4.(1)_{-2}^{+2}$   &   $4.(1)_{-3}^{+3}$   &   $4.(1)_{-1}^{+1}$\\ 
		$\Omega \rm (rad)$   &   $5_{-1}^{+1} \times 10^{-2}$   &   $5_{-1}^{+1} \times 10^{-2}$   &   $4.(1)_{-8}^{+8} \times 10^{-2}$\\ 
		$\rm DM (cm^{-3}\, pc)$   &   $1.885(7)_{-5}^{+5} \times 10^{2}$   &   $1.885(7)_{-5}^{+5} \times 10^{2}$   &   $1.885(6)_{-5}^{+4} \times 10^{2}$\\ 
		\hline
		$r/R_*$   &   $7_{-4}^{+4}$   &   $1.7_{-1.5}^{+0.3} \times 10^{2}$   &   $1.(0)_{-5}^{+5} \times 10^{3}$\\ 
		$a/r$  \space \space | \space \space $a/R_*$    &   $4_{-1}^{+1} \times 10^{-2}$   \space \space | \space \space $0.3_{-0.2}^{+0.1}$     &   $4_{-1}^{+1} \times 10^{-2}$    \space \space | \space \space  $5_{-4}^{+2}$   &   $3.(4)_{-8}^{+8} \times 10^{-2}$   \space \space | \space \space  $33_{-16}^{+16}$ \\ 
		$b/r$ \space \space | \space \space $b/R_*$   &   $3_{-1}^{+1} \times 10^{-2}$     \space \space | \space \space $0.2_{-0.1}^{+0.1}$   &   $3_{-1}^{+1} \times 10^{-2}$    \space \space | \space \space   $4_{-3}^{+2}$   &   $3.(0)_{-8}^{+8} \times 10^{-2}$   \space \space | \space \space   $29_{-14}^{+14}$  \\ 
	\end{tabular*}
\tablefoot{Median values and median uncertainties for the three main modes of the posterior sample of source A, as well as the corresponding derived values (see Tab. \ref{tab:res}). These three modes total 97\% of the weight.}
\end{table*}

\begin{table*}[h]
	\caption{Same as Table \ref{tab:frbAmodpars} for source B. The two modes represent 100\% of the sample. \label{tab:frbBmodpars}}
	\begin{tabular*}{\textwidth}{@{\extracolsep\fill}lcccc}
		Weight   &   $0.39$   &   $0.61$\\ 
		\hline
		$\mathrm{log_{10}}(f_*/\mathrm{MHz})$   &   $3.(4)_{-2}^{+2}$   &   $4.(6)_{-4}^{+4}$\\ 
		$i \rm (rad)$   &   $1.6_{-1}^{+0.9}$   &   $1.9_{-1}^{+0.7}$\\ 
		$P_* \rm (s)$   &   $0.8_{-0.2}^{+0.1}$   &   $0.8_{-0.1}^{+0.1}$\\ 
		$\mathrm{log_{10}}(\alpha)$   &   $0.2_{-0.3}^{+0.3}$   &   $1.(0)_{-4}^{+4}$\\ 
		$p$   &   $3.(8)_{-1}^{+1}$   &   $3.(7)_{-1}^{+1}$\\ 
		$\Omega \rm (rad)$   &   $5_{-1}^{+1} \times 10^{-2}$   &   $5_{-1}^{+1} \times 10^{-2}$\\ 
		$\rm DM (cm^{-3}\, pc)$   &   $3.489(1)_{-5}^{+5} \times 10^{2}$   &   $3.489(0)_{-5}^{+5} \times 10^{2}$\\
		\hline
		$r/R_*$   &   $5_{-3}^{+2}$   &   $1.(2)_{-9}^{+9} \times 10^{2}$\\ 
		$a/r$  \space \space | \space \space $a/R_*$   &   $2.(0)_{-6}^{+6} \times 10^{-2}$ \space \space | \space \space  $0.3_{-0.2}^{+0.2}$    &   $1.(9)_{-5}^{+5} \times 10^{-2}$  \space \space | \space \space $7_{-5}^{+5}$\\ 
		$b/r$  \space \space | \space \space $b/R_*$   &   $6_{-2}^{+2} \times 10^{-2}$ \space \space | \space \space  $0.1(1)_{-7}^{+5}$   &   $6_{-2}^{+2} \times 10^{-2}$  \space \space | \space \space  $2_{-2}^{+2}$\\ 
	\end{tabular*}
\end{table*}
\FloatBarrier

\end{appendix}

\end{document}